# Beyond Epitaxy: Ion Implantation as a New Tool for Orbital Engineering

Andreas Herklotz[1,*], Jonathan R. Petrie[2], Thomas Z. Ward[3,+]

[1] Institute for Physics, Martin-Luther-University Halle-Wittenberg, Halle, Germany
[2] Materials Science and Technology Division, Oak Ridge National Laboratory, Oak Ridge, TN, USA
[3] Center for Nanophase Materials Sciences, Oak Ridge National Laboratory, Oak Ridge, TN, USA

[*] herklotza@gmail.com
[+] wardtz@ornl.gov

**Manipulating electronic orbital states in quantum materials provides a powerful means to control their physical properties and technological functionality. Here, we demonstrate that orbital populations in strongly correlated oxide thin films can be continuously and reversibly tuned through post-synthesis He ion implantation. Using $LaNiO_3$ as a model system, we show that the orbital preference can be systematically adjusted from favoring in-plane $d_{x^2-y^2}$ occupation toward out-of-plane $d_{z^2}$ states through precise control of ion fluence. X-ray linear dichroism measurements reveal this orbital reconstruction, while density functional theory calculations show the effect stems from ion implantation induced changes in unit cell tetragonality. Unlike conventional heteroepitaxial approaches that lock in orbital configurations during growth, this strain doping technique enables continuous orbital tuning and selective modification of specific film regions after device fabrication. We demonstrate the practical impact of this control by achieving a seven-fold enhancement in oxygen reduction reaction catalysis that can be reversibly tuned through the implantation process. This work establishes ion implantation as a powerful approach for orbital engineering that complements existing synthesis-based strategies while offering unique advantages for both basic research and device development.**

**Introduction**

Orbital engineering has recently been a much-pursued path to enhance functionalities in solids [1,2]. This is particularly true for transition metal oxides, where the intricate coupling between lattice, spin, charge and orbital degrees of freedom can be used to generate and control quantum phenomena [3]. Examples for the successful implementation of orbital engineering are the control of the Mott transition in $VO_2$ films [4], stabilizing 3D charge order in a cuprate superconductor [5] and the manipulation of spin-orbit torque [6]. Common approaches to control orbital occupation and order typically involve the growth of thin film heterostructures. In these heterostructures, orbital engineering is then, for example, established through strain [7], spatial confinement [8], symmetry breaking [9,10] or charge transfer [11].

Heteroepitaxial strain is the most widely used method to manipulate lattice distortions and resulting orbital affects in transition metal oxides. In $3d$ transition metal $ABO_3$ perovskites, such as the strongly correlated $LaNiO_3$ (LNO), the hybridization between $B$-site $3d$ states and O $2p$ states can provide a strong coupling mechanism between structural and orbital degrees of freedom. In a cubic perovskite unit cell, the crystal field of the $BO_6$ octahedra splits $3d$ orbitals into $e_g$ and $t_{2g}$ bands. Strain induced structural distortion of these octahedra can be used to shift orbital degeneracy. In the simplest case of a tetragonal unit cell distortion, the $e_g$ band splits into $d_{z^2}$ and $d_{x^2-y^2}$ orbitals, analogous to the well-known Jahn-Teller effect. LNO exhibits a Ni $3d^7$ electronic configuration with a half-filled $e_g$ band which can be driven toward the lower lying $e_g$ orbital with increased tetragonal distortion. Functionally, the application of compressive heteroepitaxial strain in LNO thin films has demonstrated that it is possible to induce out-of-plane orbital polarization that can generate enhancement in oxygen evolution and oxygen reduction catalytic applications due to greater orbital interactions at the reaction surface [7].

Standard thin film heterostructuring locks in a defined strain state and does not allow for post-synthesis control of the orbital occupation. However, thin film research as well as potential applications would largely benefit from the capability to fine-tune strain states and the ability to locally pattern strain states into a thin film. Strain doping has been developed as a method to address this need [12]. Here, post-synthesis low-energy

He ion implantation is used to effectively induce a uniaxial out-of-plane lattice expansion. Previous work has demonstrated that strain doping can be applied to a large variety of materials where strain effects dominate over defect induced phenomena [13–16]. The uniaxial nature of strain doping makes it an ideal tool to manipulate the unit cell tetragonality of a film in a highly tunable and selective way that enables direct study of strain dependence on orbital occupation and resulting functional properties. Prior works on $La_{0.7}Sr_{0.3}MnO_3$ thin films used theoretical models to suggest uniaxial lattice expansion through strain doping induced changes to orbital occupation which could be manipulated to shift phase transition temperatures and electronic phases [12]. However, experimental evidence of orbital states across implantations was not provided. Likewise, Wang *et al.* found a metal-insulator-transition in He ion implanted LNO films, but did not measure or correlate these results with orbital polarization [17].

In this work we apply strain doping to epitaxial LNO films. Using X-ray spectroscopy we provide direct proof that uniaxial strain induced by ion implantation changes the orbital occupancy of Ni $3d$ $e_g$ states. The results are in good qualitative agreement with density functional theory. Comparing the effects of strain doping on uniaxial out-of-plane lattice expansion with as-grown heteroepitaxial biaxial in-plane strain demonstrates that unit cell tetragonality can be manipulated and used to shift the orbital polarization of the nickelate film. As a functional demonstration, oxygen reduction reaction measurements are conducted which show that the change in orbital polarization leads to strongly enhanced catalytic activity, surpassing the improvements that can be achieved by epitaxial in-plane strain alone.

**Results**

Three epitaxial (001)-oriented LNO films (12 nm thickness) were deposited by pulsed laser deposition on $TiO_2$-terminated $SrTiO_3$ (001) substrates. Two of these films were then implanted post-synthesis with 4 keV He ions and a fluence of $5 \times 10^{15}$ and $10 \times 10^{15}$ He/cm$^2$, respectively. As seen by the presence of thickness fringes in X-ray diffraction (XRD) *θ-2θ* scans in **Figure 1a**, the as-grown films are deposited with a high structural film quality. The reciprocal space map in **Figure 1b** demonstrates that as-grown films are in-plane coherently strained to the lattice of the substrate. The in-plane strain is about

1.7%. The elastic reaction of LNO leads to an out-of-plane strain of -1.2% compared to the bulk lattice. The consequent pseudocubic unit cell tetragonality $t = c/a$ is 0.971.

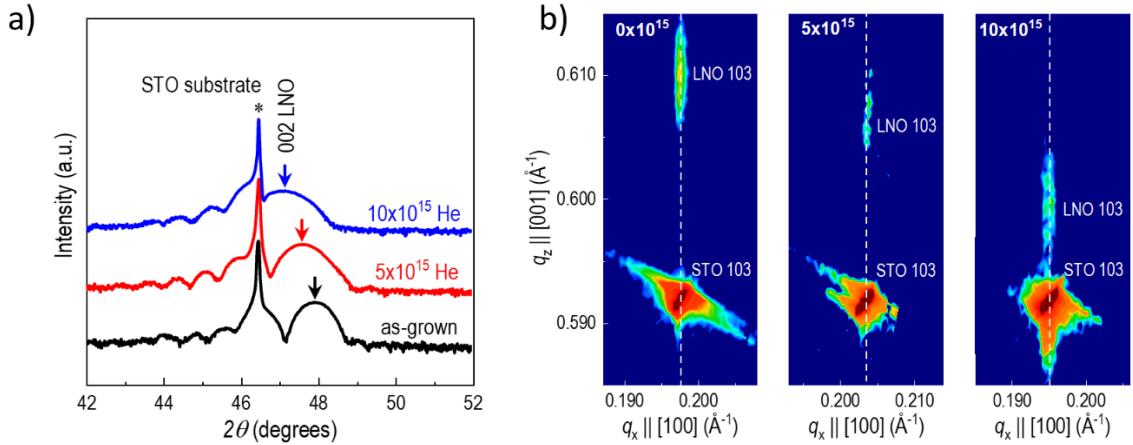

**Figure 1.** a) XRD $\theta$-$2\theta$ scan of the $(002)_{pc}$ peak of 12 nm LNO films on STO implanted with 0 to $10\times10^{15}$ He/cm$^2$. b) Reciprocal space maps around the $(103)_{pc}$ LNO peak for the three samples.

Under He ion implantation, the (002) XRD peak of the LNO film shifts towards lower $2\theta$ values. This shift is a result of the out-of-plane lattice expansion induced by strain doping. The reciprocal space maps demonstrate that during this elongation the in-plane parameters of the films remain fixed to the substrates, i.e. the lattice expansion is uniaxial and directed only along the out-of-plane direction. For the $10\times10^{15}$ He/cm$^2$ dosed film, the uniaxial expansion compared to the as-grown film is 1.6%. This also means that the unit cell tetragonality is enhanced from 0.971 to a nearly cubic 0.987. We note that the thickness fringes remain largely intact. This fact suggests that defect creation during the ion implantation process is rather small and that changes to the thin film properties can be mainly attributed to strain effects.

Linear dichroism through X-ray absorption spectroscopy (XAS) was used to characterize the evolution of orbital occupation under ion implantation. **Figure 2a** shows absorption spectra on the Ni-L$_2$ edge for both, X-rays polarized perpendicular ($E//c$) and parallel ($E//ab$) to the film plane. These spectra probe the energies and unoccupied states of the $d_{z^2}$ and $d_{x^2-y^2}$ orbitals, respectively. We note that for all three films there is no shift in the polarization-averaged edge, indicating stoichiometric Ni$^{3+}$ and no increase in observable oxygen defect generation under ion irradiation. For the as-grown film, the peak position as well as intensity of the $E//ab$ curve is smaller than that of $E//c$. This observation is in agreement with previous work on the effect of epitaxial strain and is a result of orbital polarization shifting towards the $d_{x2-y2}$ orbital under tensile in-plane strain [7,18] and other nickelates [19].

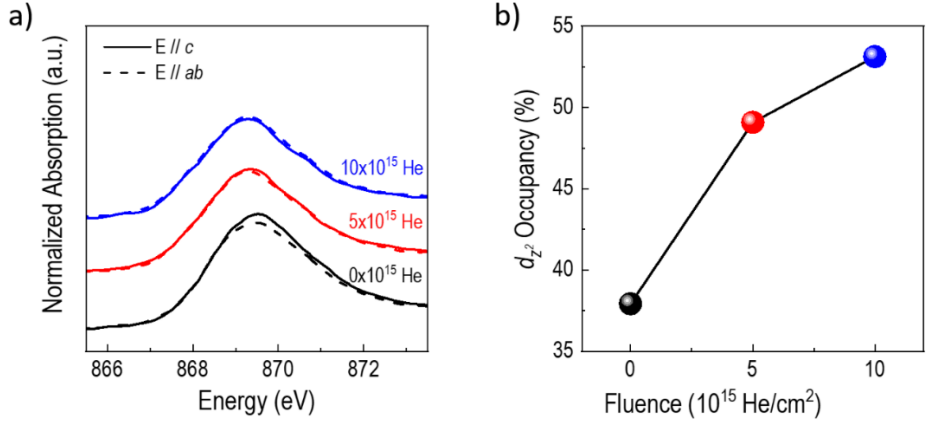

**Figure 2.** a) XAS spectra of the Ni L2 peak for LNO films on STO after increasing the out-of-plane lattice parameter with increased He implantation. The absorption along the *c*-axis (*ab*-axis) corresponds to holes in the $d_{z2}$ ($d_{x2-y2}$) orbital. b) Occupation of the $d_{z2}$ determined by integration of XAS spectra and the application of sum rules. A trend towards enhanced $d_{z2}$ orbital occupation is seen with increasing ion dose.

We find that uniaxial out-of-plane strain reverses this trend. For the $5\times10^{15}$ He/cm$^2$ film both absorption spectra are almost identical, indicating a nearly vanishing difference in orbital occupation. The *E//c* spectra of the $10\times10^{15}$ He/cm$^2$ film has a slightly smaller peak position and intensity compared to the *E//ab* spectra. The orbital polarization is shifted towards the $d_{z2}$ orbital with uniaxial lattice expansion. **Figure 2b** shows the $d_{z2}$ orbital occupancy as function of He ion dose as calculated using sum rules [10]. The $d_{z2}$ occupancy is effectively shifted from about 36 to 53%. Strain doping countereffects the depopulation of the $d_{z2}$ orbital induced by tensile biaxial strain by increasing unit cell volume along the out-of-plane direction.

To further isolate the effect of uniaxial strain we have performed density functional theory (DFT) calculations of LNO under biaxial in-plane and uniaxial out-of-plane strain. **Figure 3a** shows the partial orbital-resolved density of states (PDOS) near the fermi energy for the $d_{x2-y2}$ (blue) and the $d_{z2}$ (red) orbitals. The PDOS spectra are shown for three different strain states. All three structures are strained in-plane by 2% and then superimposed with uniaxial out-of-plane strain of 0%, 2% and 4%, respectively. This procedure mimics the effects of strain doping of a film coherently grown on $SrTiO_3$ substrates. The as-grown LNO model has a higher $d_{z2}$ PDOS slightly below the fermi level, but a lower PDOS above the fermi level. However, when uniaxial strain is imposed, the difference between the $d_{z2}$ and $d_{x2-y2}$ PDOS decreases and practically vanishes under 4% out-of-plane strain. This theoretical result suggests that uniaxial strain has the opposite effect to tensile biaxial in-plane strain and is fully consistent with the experimental observations.

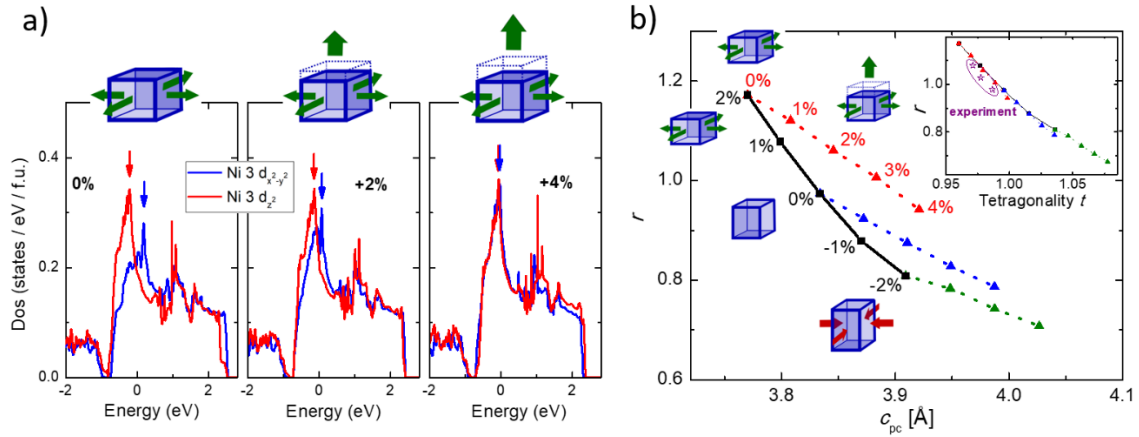

**Figure 3.** a) Orbital-density of states for the two Ni 3d $e_g$ orbitals under a fixed tensile in-plane strain of 2%, but different uniaxial out-of-plane strain. b) Calculated hole ratio $r$ under biaxial in-plane strain (black solid line) and uniaxial out-of-plane strain (colored dotted lines) as a function the pseudocubic out-of-plane lattice parameter. The inset shows the same data as function of the unit cell tetragonality. The experimental data was added to this figure.

A quantity that can be determined by both, DFT and XAS measurements, is the hole ratio $r = h_{z2} / h_{x2-y2}$, where $h_j$ is the number of holes in the orbital $j$ [10]. In **Figure 3b** we plot the theoretically calculated $r$ as the function of the pseudocubic LNO out-of-plane parameter for different strain scenarios. The black solid line presents the result for compressive and tensile biaxial in-plane strain. A large tunability of the hole ratio is observed. For tensile strain the hole ratio is above one, meaning that the Ni $3d_{z2}$ orbital is occupied with more holes than the $d_{x2-y2}$ orbital, i.e. the $d_{z2}$ has an orbital occupation below 50%. Uniaxial out-of-plane strain, represented by dotted lines, lowers the hole ratio. This lowering happens regardless of the as-grown biaxial strain state of +2% (red), 0% (blue) and -2% (green). Interestingly, this predicts that it is likely possible to apply ion implantation to compressively grown thin films to push orbital polarizations into regimes currently inaccessible to heteroepitaxy strain.

In the LNO/SrTiO$_3$ case, uniaxial strain through He ion implantation is counteracting the effect of tensile biaxial in-plane strain. The main reason for this observation is that uniaxial strain increases the tetragonality of the LNO unit cell, and in this scenario, pushes the film back towards a cubic state. The inset of **Figure 3b** shows the hole ratio *r* as determined from DFT calculations as function of the unit cell tetragonality. A nearly linear relationship is observed, highlighting that the tetragonality of the perovskite unit cell is the dominant factor for tuning orbital polarization. Using the in-plane and out-of-plane lattice parameters determined from the XRD measurements show good agreement between experimental and theoretical values. This result further corroborates that He ion implantation is driving the orbital occupation of the nickelate through uniaxial strain, rather than defect-related phenomena.

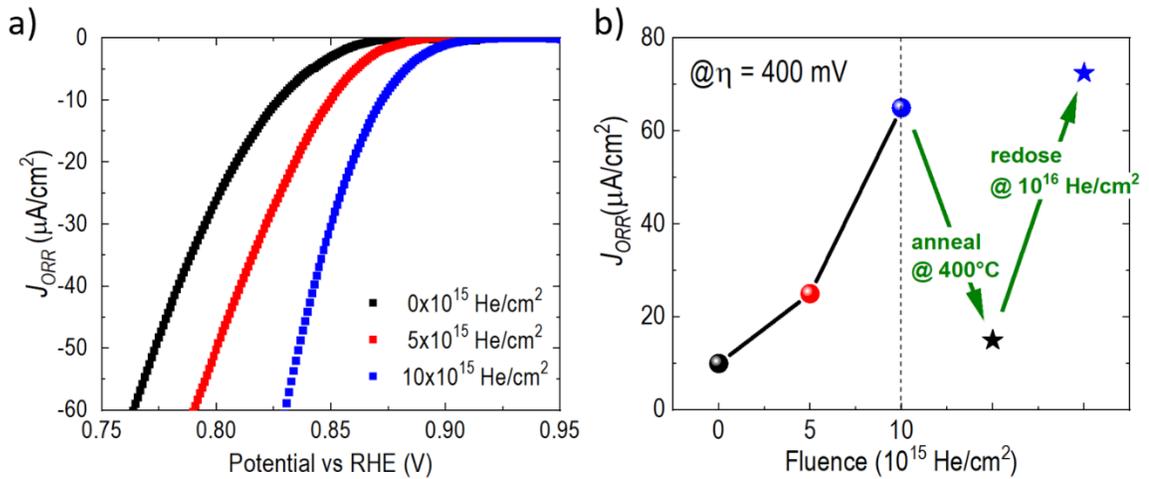

**Figure 4.** a) Polarization curves for the ORR for films implanted with different doses of Helium. b) Current density at an overpotential η of 400 mV (ORR = 0.823 V). After implantation with $10^{15}$ He/cm$^2$ the film was annealed for 12h at 400°C, remeasured (black star), redosed with $10^{15}$ He/cm$^2$ and consequently remeasured again (blue star).

It was previously demonstrated that controlling orbital populations through heteroepitaxial strain is a powerful tool to improve the performance of catalytic active materials [20], including LNO films [7]. Strain doping may thus be expected to provide a novel alternative means to finely tailor catalytic activity. **Figure 4a** shows a measurement of the oxygen reduction reaction (ORR) as a function of He ion fluence. We find a strong enhancement of the current densities with increasing fluence. This increase is in qualitative agreement with the previous study on biaxial in-plane strain, where in-plane compression and the consequent enhanced unit cell tetragonality lead to a significantly enhanced ORR reaction as a result of increased $d_{z^2}$ occupancy [7]. As expected, the as-grown LNO film's ORR response is measured to be nearly identical to that reported for the LNO film grown on STO. However, while the previous work demonstrated an increase in reactivity by inducing static tetragonality using heteroepitaxial compressive strain induced by a $LaAlO_3$ substrate, the present work shows that increases in reactivity can be induced in a continuous manner on the tensile grown film without the need to change substrate. Not only does this trend support the previous work's assertion that increasing out-of-plane orbital populations is a means of increasing ORR, it also suggests that moving toward an artificially large unit cell volume not limited by traditional Poisson effect may allow further enhancements to reactivity. Not only is there a clear increase in ORR with increasing uniaxial lattice expansion, but the ORR measured at the highest strain doping boosts the current density at an overpotential of 400 mV by almost a factor of 7 from ~10 $\mu A/cm^2$ to ~65 $\mu A/cm^2$ (**Figure 4b**) which is nearly twice the relative measured change achieved by epitaxy alone [7]. Interestingly, the strain doping effect is almost fully reversible when the He atoms are thermally annealed out of the LNO film in 400°C (**Figure 4b**). It is worth noting that this example is given to provide evidence of the efficacy of strain doping in controlling a well-documented orbital induced functional response. A more thorough study is required to fully quantify the limits of catalytic control through strain doping. However, the importance of this result is that large improvements in catalytic activity can be created post-synthesis in a room temperature process that is compatible with virtually all thin films preparation technologies.

## Conclusion

We have demonstrated that uniaxial lattice expansion induced by strain doping is a highly effective method to tune orbital polarization in LNO films. The successful application of strain doping to other material systems and the general nature of strain-induced orbital splitting suggests that this approach can be used on a wide range of transition metal oxides. The ability to fine-tune orbital occupancy through the ion implantation dose is particularly interesting in cases where lifting or restoring orbital degeneracy plays an essential role and will be of fundamental importance to understanding and controlling in other complex systems such as cuprate or iron-based superconductors and topological insulators [21,22]. Importantly, ion implantation can be conducted locally through masking or direct focused ion irradiation which opens many possibilities for monolithic device fabrication of emerging electronic and sensing applications.

**Methods Summary**

*Heterostructure growth*: Epitaxial LNO films (12 nm in thickness) were grown on various oxide substrates by pulsed laser deposition. The LNO growth temperature, oxygen partial pressure, laser fluence, and repetition rate were optimized at 600ºC, 100 mTorr, 1.5 J/cm$^2$, and 10 Hz, respectively.

*He ion implantation*: After film growth, Au films of 15 nm thickness have been deposited on top of the sample to serve as a buffer and neutralization layer for helium ion implantation. The sample was cut into smaller pieces and various Helium doses were implanted using a *SPECS IQE 11/35* ion source at an energy of 4 keV. After implantation, the Au layers were mechanically removed.

*X-ray diffraction*: X-ray diffraction was carried out using a *Panalytical X'Pert* thin film diffractometer with Cu $K_\alpha$ radiation.

*XAS*: XAS measurements were performed at the beamline 4-ID-C of the Advanced Photon Source at Argonne. National Laboratory. The measurement data presented was taken in fluorescence mode.

*Density functional theory calculations*: DFT calculations were performed using the Perdew-BurkeErzenhoff (PBE) exchange-correlation functional and ultrasoft potentials as implemented in the Quantum ESPRESSO (PWSCF v.6.4.1). A planewave cutoff of 790 eV and a Hubbard U value of 4 eV on the Ni d-states were employed for all calculations. Epitaxial strain was imposed by constructing a supercell of $\sqrt{2}x\sqrt{2}x2$ pseudocubic unit cells with bulk-like rhombohedral structure and fixed the in-plane strain to the desired strain. Internal structural parameters are than relaxed. In order to impose uniaxial strain, the out-of-plane lattice vector was elongated and internal structural positions are again relaxed. For self-consistent calculations a $6 \times 6 \times 6$ k-point was employed, while for band structure calculations a 13x13x13 mesh was used.

*Electrochemical characterization*: The ORR characterization was performed at 25 °C in a 150 ml solution of $O_2$-saturated 0.1 M KOH developed with Sigma-Aldrich KOH pellets and Milli-Q water. A three-electrode rotating disk electrode (RDE) setup was used with a Pt counter electrode and standard calomel (SCE) reference electrode. Potential was applied via a Biologic SP-200 Potentiostat at 5 mV/s and the samples had a rotating speed of 1600 rpm. Ohmic losses due to the film and solution were determined via a high frequency (~100 kHz) impedance measurement and subtracted from the applied potential to obtain iR-corrected currents. Before determining polarization curves, the potential was cycled at least 50 times at a scan rate of 50 mV/s between –0.3 and 0.7 V vs SCE to expose a stable surface under these ORR/OER conditions. Subsequent polarization curves were taken at a scan rate of 5 mV/s at 1600 rpm at least three times to ensure reproducibility. More details can be found in the supplemental material of Ref. [7].

### Acknowledgements

This material was based on work supported by the U.S. DOE, Office of Science, Basic Energy Sciences, Materials Science and Engineering Division at Oak Ridge National Laboratory (synthesis and characterization). Some characterization work was conducted as part of a user project at the Center for Nanophase Materials Sciences (CNMS), which is a US Department of Energy, Office of Science User Facility at Oak Ridge National Laboratory. AH was funded by the German Research Foundation (DFG) - Grant No. HE8737/1-1.

### Additional information

Correspondence and requests for materials should be addressed to herklotza@gmail.com.

### Competing financial interests

The authors declare no competing financial interests.